\documentclass[12pt,preprint]{aastex}

\slugcomment{accepted for publication in the AJ}

\shorttitle{Post-AGB Binary HD~46703}

\shortauthors{Hrivnak et al.}

\begin{document}

\title{A Spectroscopic and Photometric Study of the Metal-Poor, Pulsating, Post-AGB Binary HD~46703}

\author{Bruce J. Hrivnak\altaffilmark{1,2}, Hans Van Winckel\altaffilmark{3}, Maarten Reyniers\altaffilmark{3,4}, David Bohlender\altaffilmark{5}, Christoffel Waelkens\altaffilmark{3},  Wenxian Lu\altaffilmark{1}}

\altaffiltext{1}{Department of Physics and Astronomy, Valparaiso University, Valparaiso, IN 46383; Bruce.Hrivnak@valpo.edu, Wen.Lu@valpo.edu}
\altaffiltext{2}{Guest Investigator, Dominion Astrophysical
Observatory, Herzberg Institute of Astrophysics, National Research
Council of Canada} 
\altaffiltext{3}{Instituut voor Sterrenkunde, K.U.Leuven University, 3001
Leuven (Heverlee) Belgium; Hans.VanWinckel@ster.kuleuven.be}
\altaffiltext{4}{Now at KMI, Department Waarnemingen, Ringlaan 3, 1180 Brussels, Belgium}
\altaffiltext{5}{Dominion Astrophysical
Observatory, Herzberg Institute of Astrophysics, National Research
Council of Canada; David.Bohlender@nrc-cnrc.gc.ca} 


\begin{abstract}
The metal-poor post-AGB star HD~46703 is shown to be a single-line
spectroscopic binary with a period of 600 days, a high velocity of $-$94 km~s$^{-1}$,
and an orbital
eccentricity of 0.3.  Light curve studies show that it also pulsates
with a period of 29 days. High$-$resolution, high signal-to-noise spectra were
used for a new abundance study.  The atmospheric model determined is
T$_{eff}$ = 6250 K, log $\it g$ = 1.0, V$_t$ = 3.0 km~s$^{-1}$, and
a metal abundance of [M/H] = $-$1.5.  A low carbon abundance and
lack of s-process element enhancement indicate that the star has not
experienced third dredge-up on the AGB.  The sulfur and zinc
abundances are high compared with iron, and the chemical abundances
show a clear anti$-$correlation with condensation temperature.  The
abundance depletion pattern is similar to that seen in other
post$-$AGB binaries, and, like them, is attributed to the chemical
fractionation of refractory elements onto dust stored in a
circumbinary disk and the re-accretion of volatiles in the stellar atmosphere.  
The infrared excess is small but the excess
energy distribution is very similar to what can expected from a
disk. HD~46703 joins the growing list of depleted, post-AGB stars
which are likely surrounded by a dusty and stable circumbinary disk.

\end{abstract}




\section{INTRODUCTION}
During the past decennia, it has been realized that post-AGB stars are
chemically much more diverse than anticipated. In the canonical
picture, the photospheric content of a post-AGB star is the
end-product of the chemical enrichment induced by the thermal pulses
of AGB stars, and as such post-AGB stars are ideal tracers for the 3rd
dredge-up nucleosynthesis. Some objects are the most s-process
enriched objects known to date \citep[e.g.,][and references therein]{vanwin03}, 
while others are not enriched at all. This
dichotomy is very strict, in the sense that mildly enhanced objects
are not often observed  \citep[and references therein]{vanwin03}.
For the moment it is not clear whether this dichotomy is caused by a selection effect,
or if it is the result of two different evolutionary channels. 

In this study, we focus on HD~46703. More than 20 years ago,
Bond and Luck drew particular attention to HD~46703 with their
abundance studies \citep{luc84,bon87}.  They found the object to be
iron-poor, [Fe/H] = $-$1.6, but relatively overabundant in carbon ([C/Fe]=+1.0),
nitrogen ([N/Fe]=+1.8), and oxygen ([O/Fe]=+1.1) and underabundant in
s-process elements (Y,Zr,Ba; ([s-process/Fe]=$-$0.6 to $-$0.8).  This
combination is hard to explain.  The atmosphere model they determined
was T$_{\rm eff}$ = 6000$\pm$150 K, log ${\it g}$ = 0.4$\pm$0.4, and
V$_{\rm turb}$ = 3.5$\pm$0.5 km~s$^{-1}$.  They also found that it was a
high velocity star, with V$_{\rm r}$ = $-$106 km~s$^{-1}$.  On this
basis, they concluded that HD~46703 was a low-mass, post-AGB star.
Surprisingly, they also found a strong overabundance of sulfur, [S/Fe]
= +1.2, even though the other iron-peak elements had similar
abundances to iron \citep{bon87}.

HD~46703 was also shown to be a light variable, with variations
peak-to-peak of 0.22 mag over a 12 year period
\citep[1966$-$1979;][]{luc84} and 0.1 mag over three 1$-$2 week observing
intervals 1981$-$1982 \citep{bon84}.  The star was bluer when brighter.
On this basis, it was classified as a low-amplitude, semi-regular
variable and has been assigned the name V382 Aur.

\citet{wat92} included HD~46703 among a small group of 
post-AGB objects for which they proposed that the selective depletion in the
stellar atmosphere arose from the formation of a circumstellar dust
disk, in which refractory elements are tied up in the dust and then
the volatiles are selectively re-accreted in the atmosphere of the star.
Since then, atmospheric depletion patterns have been detected in many
more objects (see the recent papers by \citet{gir05, maa07,
rey07b} and references therein). Although the process itself is
still not well explored theoretically, it seems to be a very common
phenomenon among Galactic post-AGB stars as well as in their peers in
the LMC.

The original sample of four extremely depleted objects turned out to be
binaries \citep{vanwin95} in which there was strong evidence for
the presence of a circumbinary disk. Many of the objects subsequently determined 
to have the chemical depletion anomaly were found to be binaries
as well \citep[and references therein]{vanwin03}.  Using N-band
interferometry, \citet{der06} and \citet{der07} were able to resolve the disks
around some depleted objects, and the main conclusion of their analysis
was that the circumstellar disks must be very compact and stable.  It
is as yet still unclear whether all such depleted stars are 
binaries surrounded by a circumstellar disk \citep{ruy06}, or
whether there are other circumstances in which depleted photospheres
can be formed \citep{gir05,maa07}.

For large-amplitude pulsators, standard chemical studies based on
static atmospheres fall short, and \citet{rao05} and \citet{gir05} link the
detection of chemical anomalies to a not-well-documented effect in which an
anticorrelation between [X/H] and the first ionization potential is created. For
pulsational stars with a significant photometric amplitude, any
detection of a chemical anomaly should be studied in more detail and
preferably using spectra over the whole pulsation cycle.

In this study, we therefore present a study of HD\,46703 which
combines a detailed chemical study using high-resolution optical
spectroscopy with a critical discussion of our photometric and
radial velocity monitoring results. This object is an ideal one in which
to study in detail the chemical anomalies and to confront those with the pulsational
analysis, the nature of the circumstellar material, and the
binary connection of the depletion process. 
The paper is organized as follows: we first analyze the 
radial velocity to investigate binarity, then the light curves to investigate pulsation, 
and then the high-resolution spectra to determine the chemical abundance.  
Finally we discuss the evolutionary nature of the object.

\section{RADIAL VELOCITY STUDY}

The monitoring of the radial velocity was
performed at the Observatoire de Haute-Provence (OHP) from 
1989 to 1996  and at the Dominion Astrophysical Observatory
(DAO) from 1991 to 1995 .  The individual velocities are listed in
Table~\ref{tab_rv}, where we have included the heliocentric Julian
date (HJD), the heliocentric radial velocity (V$_{\rm r}$), 
and the observatory.  

\placetable{tab_rv}

Radial velocity data obtained at OHP used the CORAVEL radial
velocity meter attached to the 1-m Swiss telescope.  
CORAVEL is a spectrophotometer by which the radial velocity 
was obtained by cross-correlations of the stellar
spectra with a hardware mask built from the spectrum of the K2~III star
Arcturus \citep{bar79}.  The S/N of the cross-correlation depended
therefore also on the spectral match between the
science target and Arcturus. The zero-point of the system was regularly
assessed during the night by measuring CORAVEL radial velocity
standards. This was done at intervals of about 1-2 hours.
All the CORAVEL observations, including ours, are held in a data-base
that is maintained by the owners of the CORAVELs in the Geneva
Observatory (Switzerland).

The observations at the DAO used the 1.2-m telescope equipped with the
DAO radial-velocity spectrometer \citep[RVS;][]{mcc85,fle82} at the Coud\'{e} focus.  A
spectral mask based on the spectrum of Procyon (F5~IV-V) was used, which
contained about 340 sharp stellar lines in the wavelength interval
from 4000 to 4600 \AA\,.  This hardware mask was cross-correlated  with the 
stellar spectrum to determine the radial velocity.  A Cd-Ar comparison lamp was observed
before and after the stellar observations.  These observations were
standardized through nightly observations of several radial velocity
standard stars (10 on a full night) from the list of \citet{sca90},
which tie the observations very closely to the IAU system
\citep[see][]{sca90}.

It is known that the CORAVEL and DAO$-$RVS velocities are on very
nearly the same standard system, except for a small color term that
affects cool stars measured with CORAVEL \citep{sca90} and which is
not significant for this F star.  Therefore we combined the data in
the following analysis.  This totaled 41 OHP$-$CORAVEL velocities and
26 DAO$-$RVS velocities.  The precision of the observations is
$\pm$0.8 km~s$^{-1}$.  We also determined three additional radial
velocities from the high-resolution spectra used in Section 5 in the
abundance analysis.

\begin{figure} \figurenum{1}\label{rv_hjd}
\plotfiddle{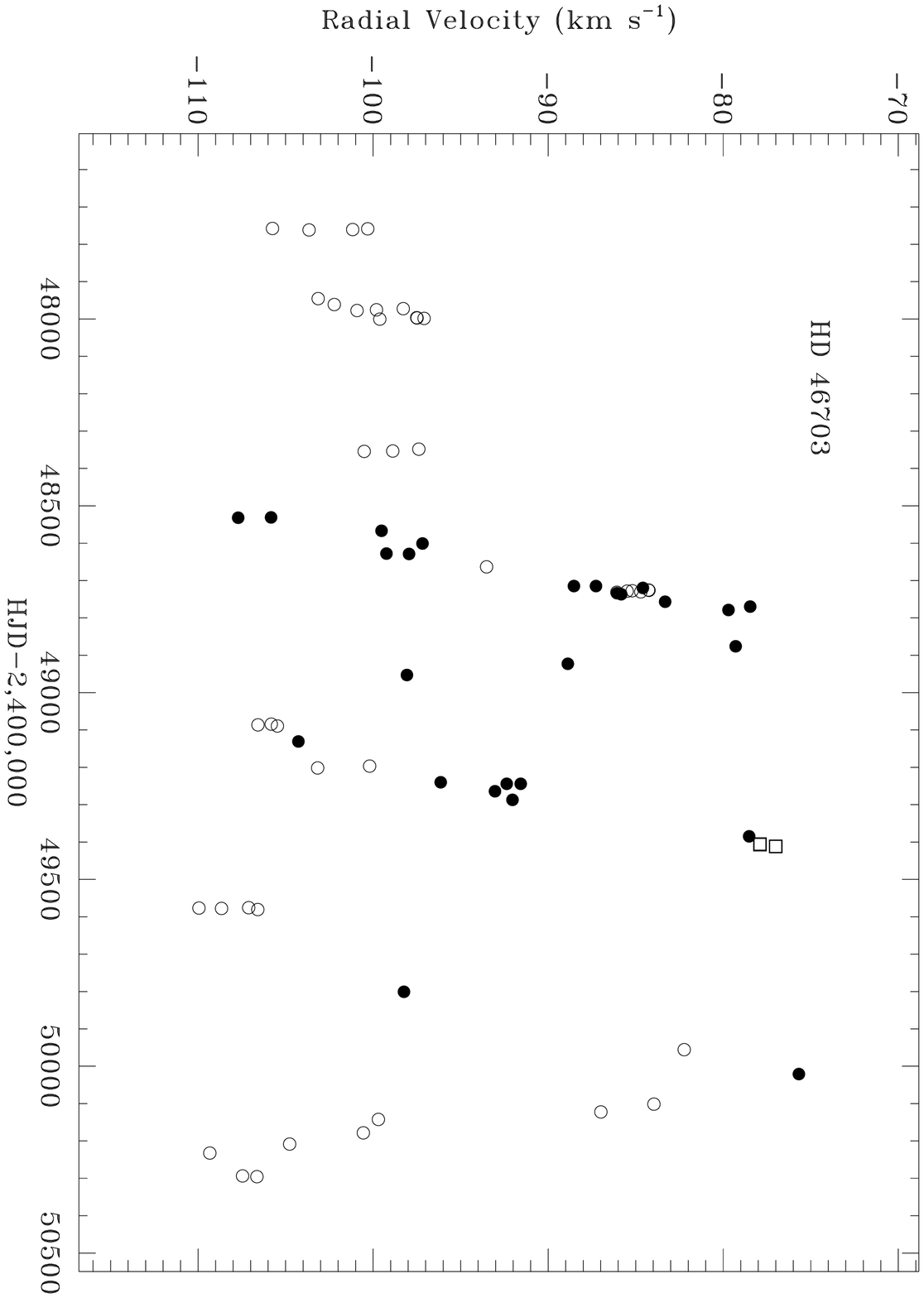}{0.0in}{+180}{400}{500}{0}{0}
\caption{Our radial velocities of HD~46703.  Open circles are the 
OHP$-$CORAVEL data, filled circles the DAO$-$RVS data, and open squares are the 
velocities derived from the high-resolution abundance spectra. }
\end{figure}

\begin{figure} \figurenum{2}\label{rv_phase} 
\plotfiddle{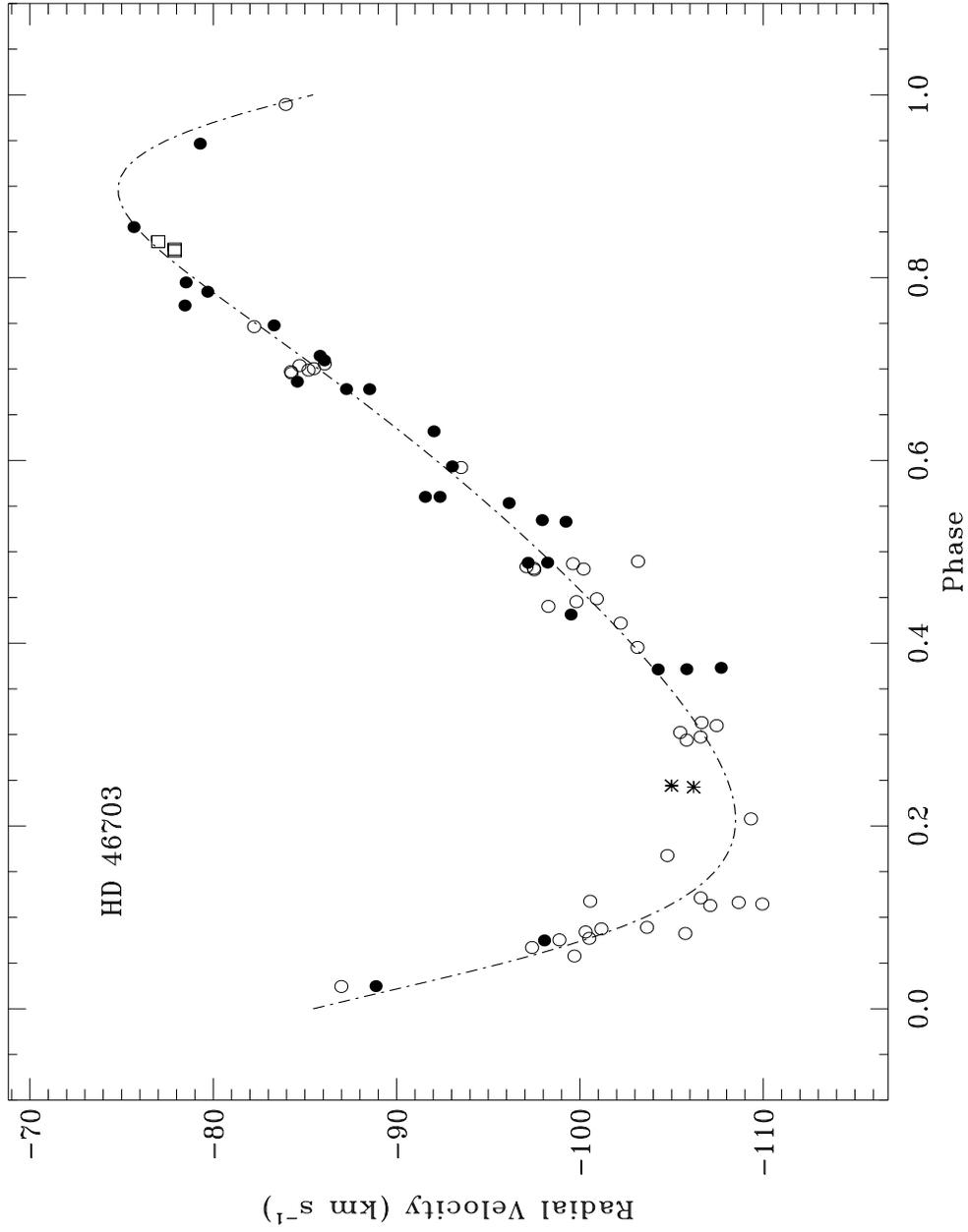}{0.0in}{+180}{400}{500}{0}{0}
\caption{Phase plot of velocities of HD~46703 fit with the binary orbit parameters of Table 2.  The symbols are the same as in Figure 1, with the additional two velocities of \citet{luc84} shown as asterisks.  }
\end{figure}

These observations show that the object clearly varies in velocity, as
shown in Figure~\ref{rv_hjd}.  The velocities range from $-$108 to
$-$76 km~s$^{-1}$.  They show a periodic pattern of about 600 days,
and were fitted with a spectroscopic binary orbit program
\citep[version from 1986]{mor75}.  We also included in the orbital
analysis the two earlier radial velocities published by \citet{luc84}.
Note that the ``discordant'' low-resolution velocity of $-$75$\pm$7
km~s$^{-1}$ from an earlier paper by \citet[see \cite{luc84}]{bon70}
is consistent with one of the extrema of the radial velocity curve.
The parameters of the resulting single-lined spectroscopic binary
orbit are listed in Table~\ref{tab_orbit}.  The system has a high
velocity of $-$93.6 km~s$^{-1}$.  The orbit is rather eccentric, with
a value of e = 0.31.  The resulting fit of the binary orbit to the radial
velocities is shown in Figure~\ref{rv_phase}.

\placefigure{rv_hjd}

\placetable{tab_orbit}

\placefigure{rv_phase}

\section{LIGHT CURVE STUDY}

Photometric observations of HD~46703 were carried out from 1986
through 2006 at three different observatories with two different
photometric systems.  These are described below.

At the Valparaiso University Observatory (VUO), photometric observations  
were carried out from 1995 through 2006 using
the 0.4-m telescope and CCD camera.  In the first several seasons the
observations were made primarily with the V filter and occasionally
with the R$_C$, but beginning in 2001 the R$_C$ filter was used
regularly.  Due to a problem that arose with the V filter, no V data
are available from the 1999$-$2000 through the 2001$-$2002 observing 
seasons.
The images
were reduced using IRAF\footnote{IRAF is
distributed by the National Optical Astronomical Observatory,
operated by the Association for Universities for Research in
Astronomy, Inc., under contract with the National Science
Foundation.}, with normal bias and flat field corrections
and using an aperture of diameter 11$\arcsec$.  Differential
magnitudes were obtained between HD~46703 and the comparison star
HD~46591 \citep[G0; the same comparison star used by][]{bon84}, for
which we measured V = 8.04 and (V$-$R$_C$) = 0.46.  This comparison
star was confirmed to be constant at the level of $\pm$0.03 mag by
comparison with two additional but much fainter stars in the field,
and our value of its V magnitude is the same as that reported by \citet{bon84}.  
The total number of
observations with each filter, and the maximum observational
uncertainty, are as follows: 84 with $\sigma\le\pm$0.007 mag in V, 48
with $\sigma\le\pm$0.004 mag in R$_C$, and 35 with $\sigma\le\pm$0.005 in
(V$-$R$_C$).  Observations of standard stars on several nights were used
to transform the differential magnitudes to the standard Johnson V and
Cousins R$_C$ systems, and in Table~\ref{tab_difmag_vuo} 
are listed the standardized differential magnitudes.
These VUO observations of HD~46703 reveal a variation in brightness over a
full range of 0.38 mag in V, with a range of almost that value in 1995$-$1996
and 2006, and a full range of 0.30 mag in R$_C$, which is seen in
2006.  The (V$-$R$_C$) colors shows a variation over a full range of
0.08 mag, reaching 0.07 in 2006, and the system is bluer when
brighter.  

\placetable{tab_difmag_vuo}

Observations were also made in the Geneva 
7-filter (U,B,V,B1,B2,V1,G)
photometric system, both with
the 0.7-m Swiss telescope at the Jungfraujoch Observatory and 
with the 1.2-m Flemish Mercator telescope at the
Roque de los Muchachos Observatory at La Palma, Spain. 
Both telescopes are equipped with a dual-channel
photometer that monitored the star and sky quasi-simultaneously, using a filter wheel
that cycled through all seven filters
four times per second. The photometer at Jungfraujoch is operated manually.  
The typical uncertainty of an observation is $\pm$0.005 mag.  
HD 46703 was observed during two major
observing sessions, which resulted in 65 
high quality measurements between 1986 Nov and 1996 Jan 
(3358 d) and 42 measurements between 2001 Oct and 2004 Mar
(884 d).  These data are listed
in Table~\ref{tab_mag_geneva}, and they 
show a peak-to-peak variation of 0.40 mag in the
V-band with a standard deviation of 0.068 mag.

\placetable{tab_mag_geneva}

\begin{figure} \figurenum{3}\label{lc_combined_hjd}  \epsscale{0.95}
\plotfiddle{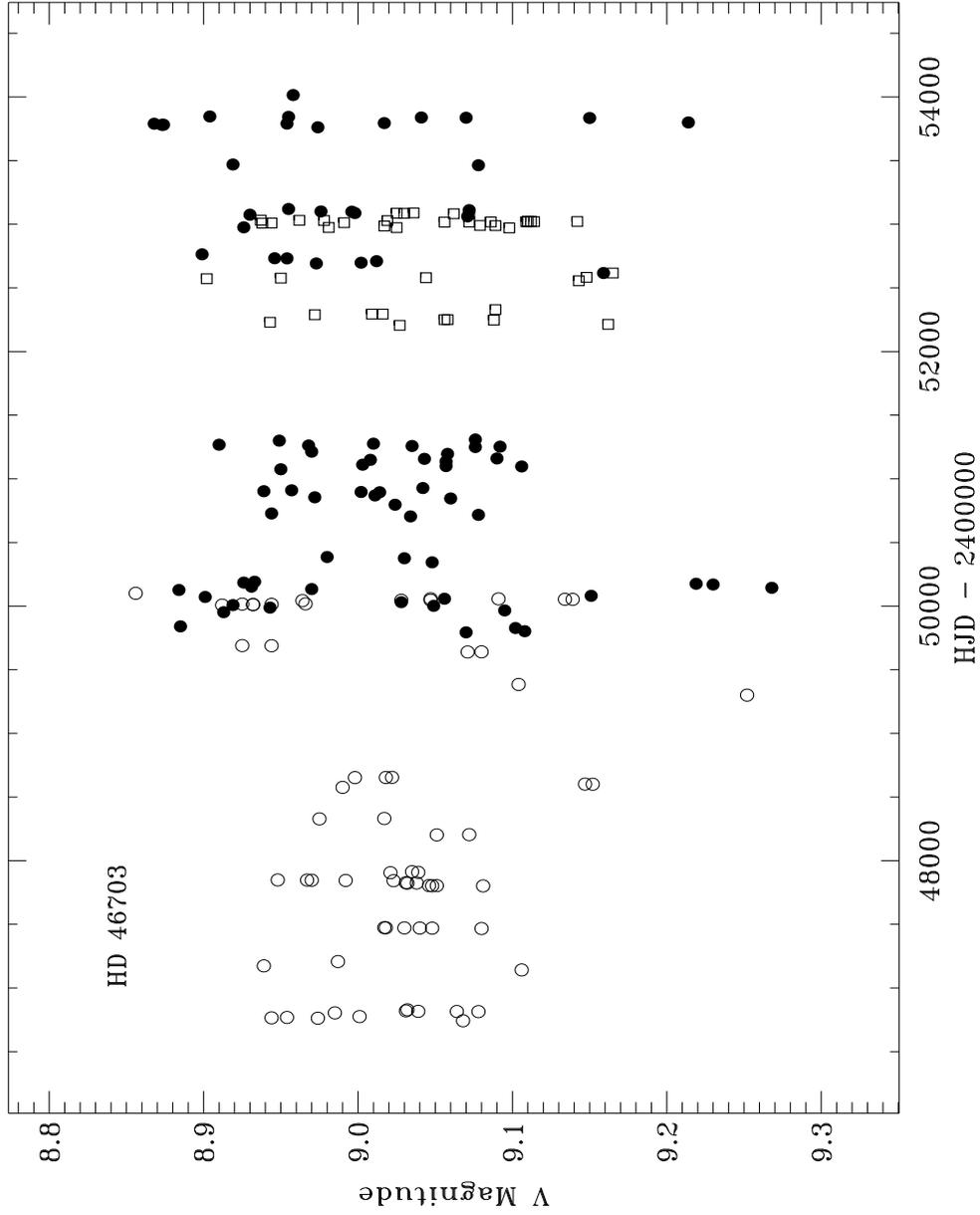}{0.0in}{+180}{400}{500}{0}{0}
\caption{V light curve HD~46703 from 1986 to 2006.  The symbols are as follows: open circle - 
Swiss telescope at Jungfraujoch, filled circle - VUO telescope, open square $-$ Mercator telescope at La Palma. } 
\end{figure}

\begin{figure} \figurenum{4}\label{color} 
\plotone{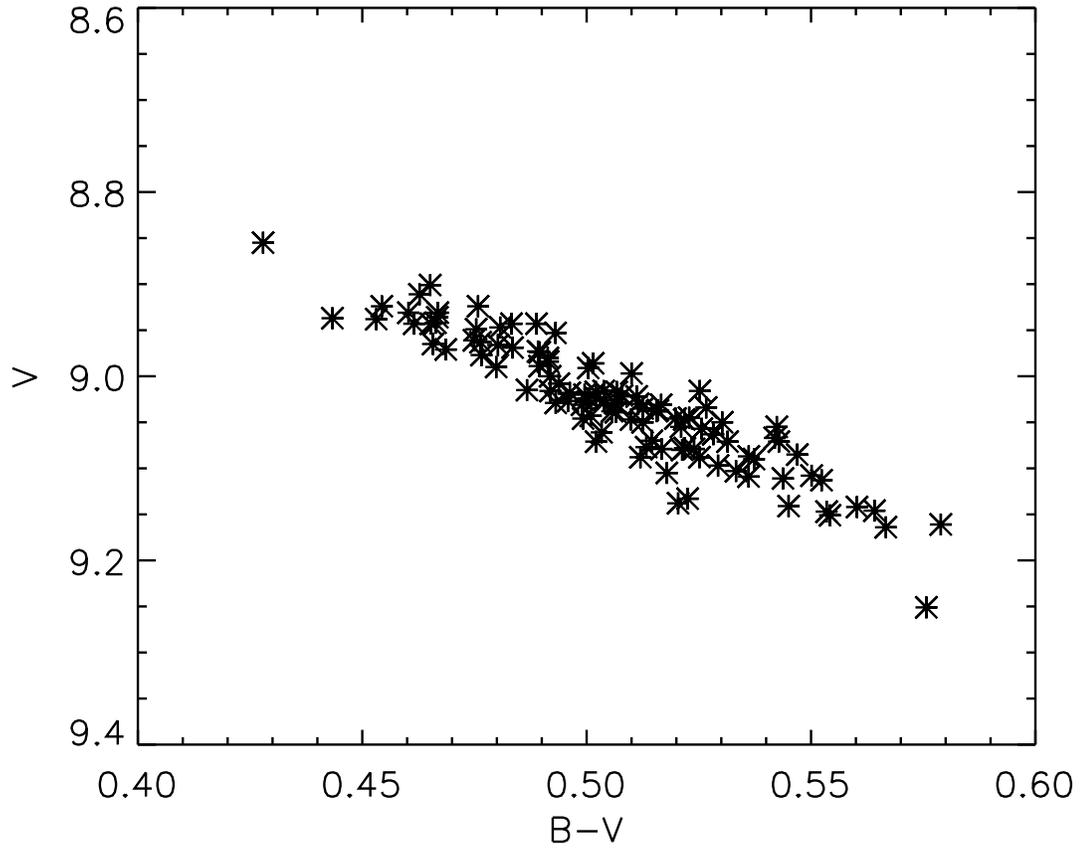}
\caption{Color-magnitude plot for HD~46703, based on the Geneva photometry transformed to the Johnson system. The system is bluer when brighter. }
\end{figure}

Since the Geneva {\it V} system is essentially identical to the Johnson {\it V} system ($\Delta$$<$0.001 mag), all of the 
{\it V} data were combined into one set and these are displayed in Figure~\ref{lc_combined_hjd}.  The data sets show good agreement.  
Several of the seasons show a clear cyclical pattern,
but even within these seasons the amplitude of the variation is not constant.
The color variations are typical of pulsations and, as noted previously, 
the object is bluer when brighter (Figure~\ref{color}).

\placefigure{lc_combined_hjd}

\placefigure{color}

Orbital light variations were investigated with the use of the combined V data set.  
These data were phased with the ephemeris determined for the 600 d orbital period. 
No significant orbital phase dependence is seen, although, as mentioned above, the light of the system does vary.

\section{PULSATIONAL VARIABILITY}

The VUO and Geneva photometric data were initially investigated
separately for variability, using the CLEAN algorithm based on a Fourier
transform \citep{rob87}.  The VUO V data suggest a possible period
of 29 d and the Geneva V data suggest possible periods of 31 and 27 d,
but none of these is very dominant.  The analysis of the combined V data 
yielded   
two closely spaced periods of 29.06 d and 28.85 d, with the former the stronger.
We further investigated the variability using the
period analysis program Period04 \citep{lenz05}.  This indicates a dominant period
of 29.1 d, with the suggestion of amplitude modulation.  We also
searched for period variability by dividing the data into four separate 
intervals of 2000 d each.  For the second interval, we see a dominant
period of 29.1 d, with an amplitude of 0.115 mag.  For the third and
fourth intervals we find a dominant period of 31.0 d with amplitudes
of 0.068 and 0.074 mag, respectively.  For the first interval the
strongest peak was about 98 d, with an amplitude of 0.054 mag; there
was a peak at about 30 d but it was not as strong.  

From the complete analysis, we conclude that the system varies with a
period of 29.0 d and a varying amplitude, with perhaps a second period
of 31 d.  
Post-AGB stars on
the hotter side of the population II Cepheid instability strip often
show a very complex pulsational pattern with irregular pulsations
\citep[e.g.,][]{kis07}. For the limited number of objects with
long-term monitoring data, similar, but slightly different periods
are found using datasets of different seasons. One of the best
studied examples to our knowledge is HD~56126, also a 
low-metallicity, F spectral type, post-AGB object. The spectroscopic and
photometric variations of HD 56126 appear to be consistent with a
non-regular radial pulsation where the dominant pulsation mode is
the first overtone. In the course of the pulsation, moderate shock
waves are repeatedly generated and they propagate throughout the
stellar atmosphere. They provoke a complex, asynchronous motion of
the outer layers \citep{fok01}. Similar behavior is found in
HD~46703, in which the main period deviates slightly during the
different datasets. This exemplifies the complex pulsation character
of luminous objects with a small envelope mass. 

We searched for a signature of the pulsation in the radial velocities by examining the residuals of the observations compared to the orbital solution.  While they vary over a range of 10 
km~s$^{-1}$, no periodicity is found in these residuals.

\section{CHEMICAL ABUNDANCE STUDY}

\begin{figure}  \figurenum{5} \label{spectrum}  
\rotatebox{270}{\plottwo{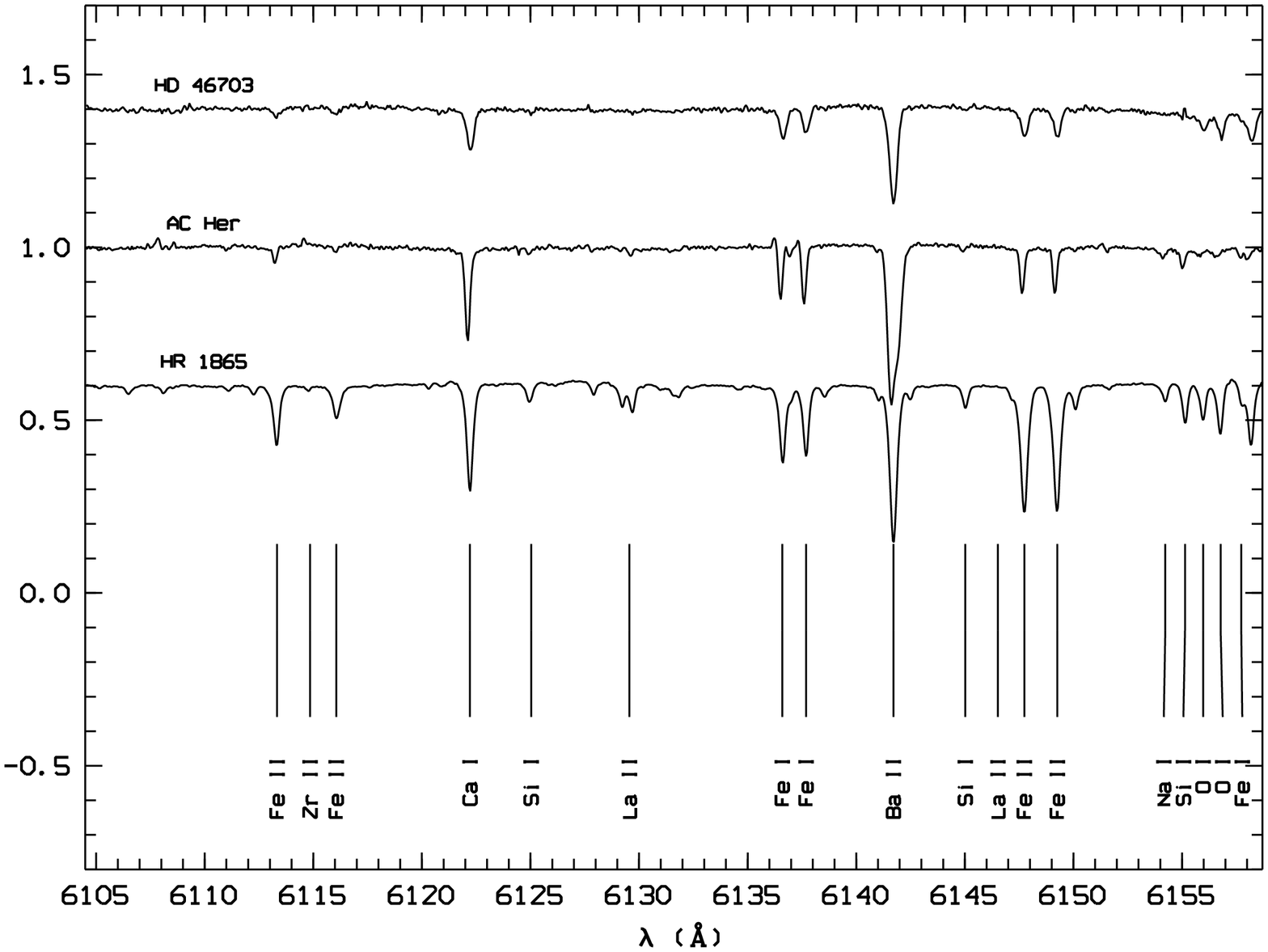}{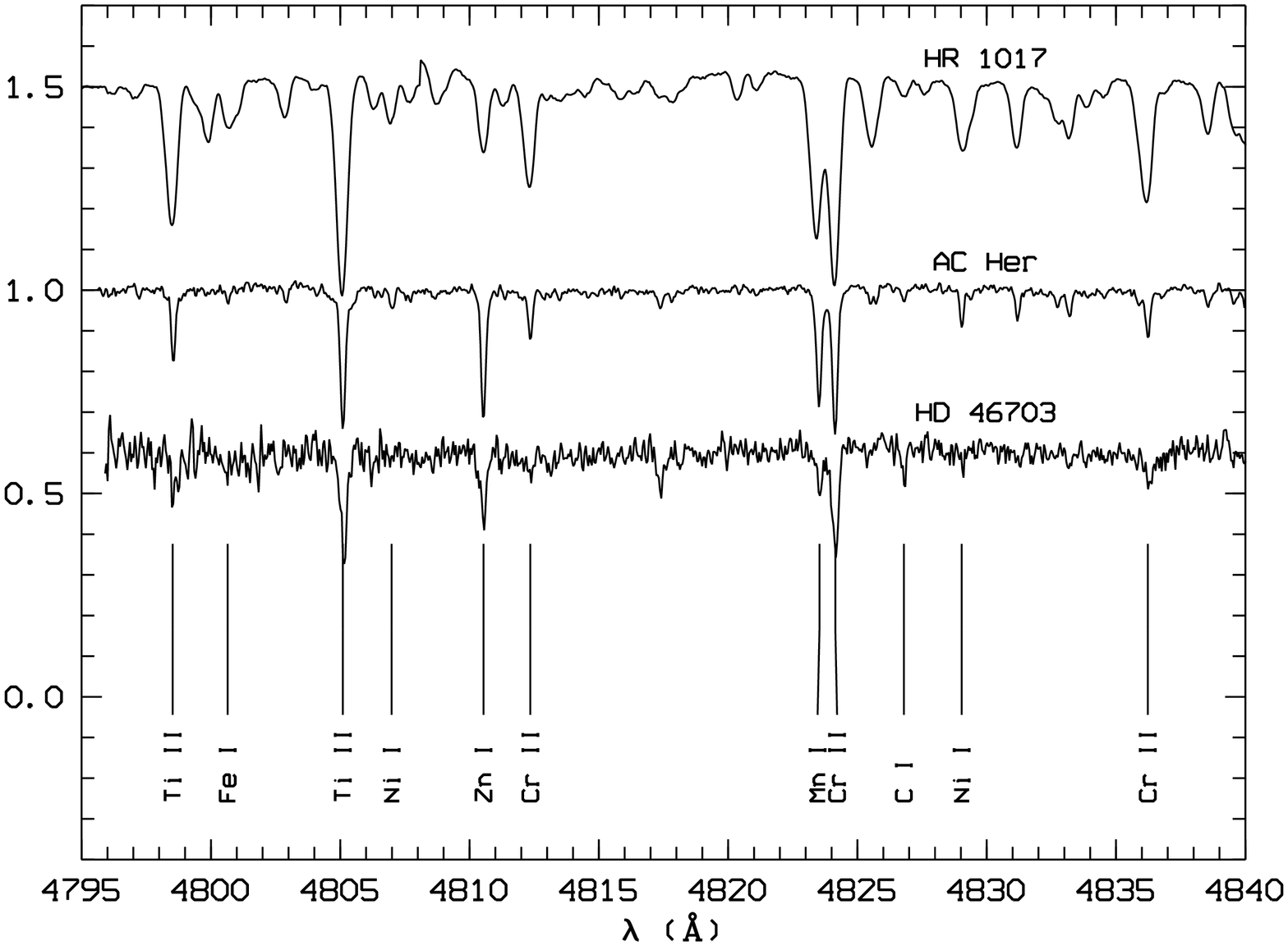}}
\caption{Plots of sample regions of the spectrum of HD~46703.  (a) Top $-$ The region around $\lambda$6150~\AA.  The weakness of the Fe lines in HD~46703 relative to those in HR~1865 (F0~Ib, solar composition) is readily seen.  (b) Bottom $-$ The region around the Zn line $\lambda$4810.54~\AA. Note the difference in strength of the Zn line relative to the other metal lines in HD~46703 compared with HR~1017 (F5~Iab, solar composition).   Also shown for comparison is the post-AGB object AC Her (metal-poor, similar depletion pattern). }
\end{figure}

High resolution, high signal-to-noise optical spectra were obtained
with the (now decommissioned) Utrecht Echelle Spectrograph (UES)
mounted on the 4.2-m William Herschel Telescope (WHT) on La Palma,
Spain. The echelle with 31.6 lines~mm$^{-1}$ was used, and the projected slit
width was 1.1$\arcsec$ on the sky, yielding a resolution of 
R\,$\simeq$\,50,000. The spectra were taken in February 1994 with a
few nights separation (Feb 21, 23 and 28), in three different settings to
cover the whole optical region (364--1022\,nm with spectral gaps from
665\,nm redwards). These UES spectra were kindly provided by Dr. Eric
Bakker and used by him in his analysis of the optical circumstellar
molecular absorption bands \citep{bak97}.
Two sample regions of the spectrum are displayed in Figure~\ref{spectrum},
with the same spectral region of a bright star (HR 1017 or HR 1865) and of
AC~Her shown for comparison. HR~1017 is an F5~Iab star (B$-$V=0.45) 
and HR~1865 is an F0~Ib star (B$-$V=0.22), 
both with solar composition, while AC\,Her (B$-$V=0.63) is a post-AGB 
object with chemical anomalies \citep{vanwin98} similar to HD~46703.
In Figure~\ref{spectrum}a is shown the spectral region around 
$\lambda6150~$\AA, displaying a number of lines of FeI, FeII, BaII, and the 
OI triplet.  The weakness of the iron lines in HD~46703 compared to those
in HR~1865 is readily apparent.
In Figure~\ref{spectrum}b is shown the spectral region around 
the Zn line $\lambda4810.54~$\AA. 
The difference in strength of the Zn line relative to the other metal lines in 
HD~46703 compared to HR~1017 can easily be seen.
The spectrum of HD~46703 has a S/N ratio of $\sim$200 in the red but is
less in the blue.

\placefigure{spectrum}

The general analysis methodology is already extensively discussed in many of our
previous papers \citep[e.g.,][]{rey04, der05}  and will not be
described in detail. We performed an LTE abundance study and determined the
fundamental parameters of the photosphere in an iterative scheme. We first
estimated the temperature (using a given gravity and metallicity) by requiring
the abundances to be independent of the excitation level. The gravity was then
fine-tuned by requiring  ionization balance for all species for which lines of
different ions could be identified. The microturbulence was quantified by
requiring the abundances to be independent of line strength. Finally, we
iterated this scheme so that convergence was found for all fundamental parameters.
We note that we used the latest ATLAS models
\citep{cas04} in combination with the latest version (April 2002) of
Sneden's LTE line analysis program MOOG \citep{sne73}. The model atmosphere
parameters were inferred from a detailed model study based on the iron lines.
The final parameters were found to be as follows: an effective temperature of 
T$_{\rm eff}$\,=\,6250\,$\pm$250 K, a gravity of $\log g$\,=\,1.0$\pm$0.05, 
a microturbulent velocity of $V_t$\,=\,3.0\,$\pm$1.0 km\,s$^{-1}$, and a model metallicity of
[M/H]\,=\,$-$1.5.

The abundance measurements derived from the individual lines of each ion are 
listed in Table~\ref{line_ew}.  
Listed for every line are the wavelength, the excitation potential ($\chi$), 
the log~{\it gf} value, the measured equivalent width (W$_{\lambda}$), the absolute
abundance ($\log\epsilon$\,$=$\,$\log$(N(X)/N(H))+12), and the abundance difference between the individual line and the mean abundance ($\Delta\log\epsilon$). 

A summary of the abundances is listed in Table~\ref{tab_abund}. The first column of
this table gives the ion, the second column gives the number of lines
used, the third one is the mean equivalent width of the lines used,
the fourth column gives mean absolute abundances derived, the fifth column is the line-to-line 
scatter ($\sigma_{\rm ltl}$) in $\log\epsilon$, and the sixth column gives the abundance
relative to the Sun [X/H]. 
The solar abundances (seventh column) are needed to 
calculate the [X/H] values \citep[for references, see][]{rey07a}. 
The dust
condensation temperatures listed (eighth column) are those given by 
\citet[][and references therein]{lod03}. They are computed using a solar
abundance mix at a pressure of 10$^{-4}$\,atm. A condensation temperature for
O is meaningless, since O is the most abundant element in rock material.

\placetable{tab_abund}

\begin{figure} \figurenum{6} 
\plotone{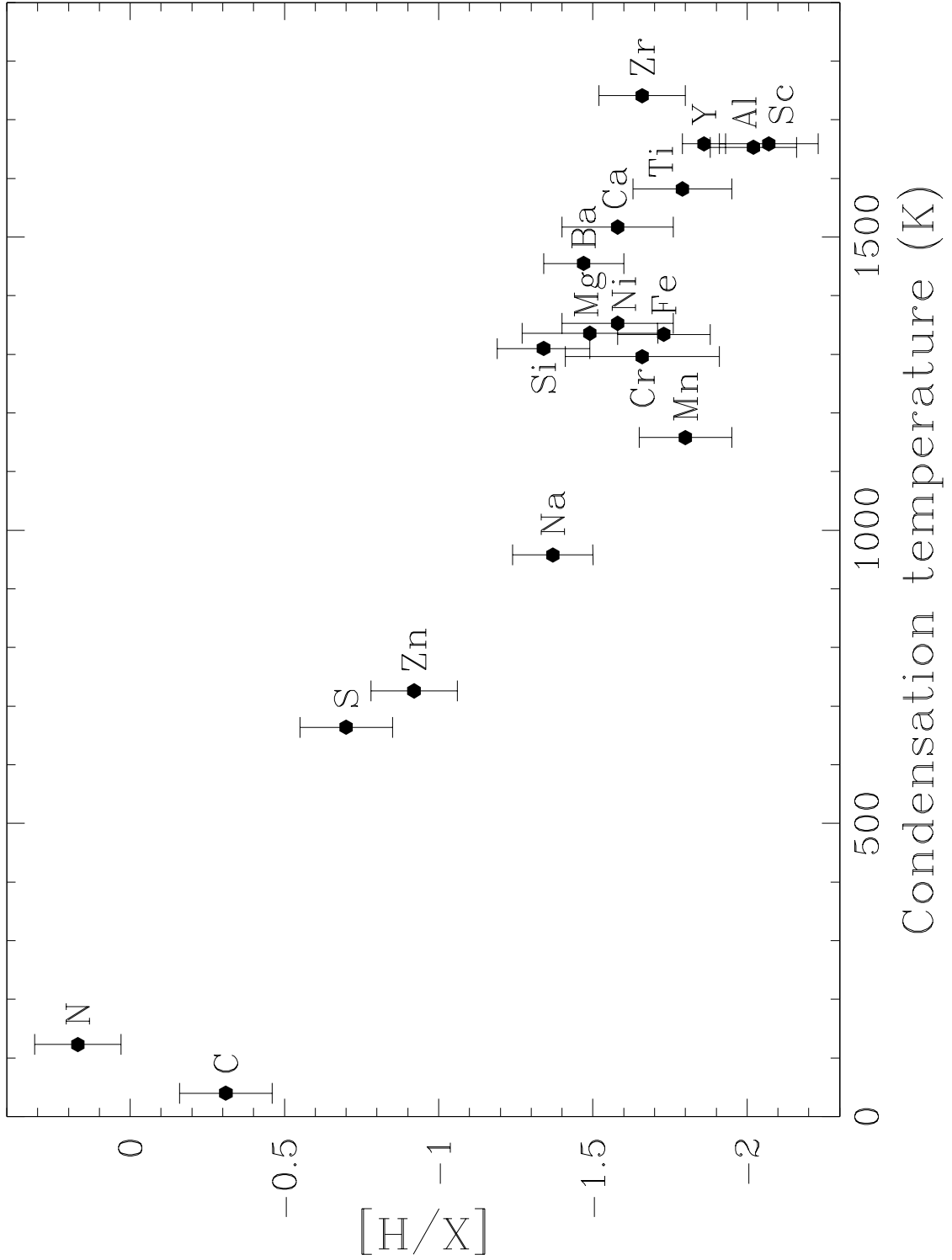}
\caption{Abundance results for HD~46703.  The anti-correlation of abundance 
with condensation temperature clearly indicates a depletion effect. \label{fig_abund}}
\end{figure}

The abundances are also presented graphically in Figure~\ref{fig_abund}, where
the abundances relative to solar [X/H] are plotted against condensation
temperature. The clear anti-correlation as seen in Figure~\ref{fig_abund} can 
immediately be recognized as a depletion pattern; elements with a low
condensation temperature, like S, Zn and the CNO-elements, have a much higher
photospheric abundance than elements with a high ($>$1000\,K) condensation temperature.
The CNO elements have low condensation temperatures and are typically expected
not to be depleted. The low carbon abundance of [C/H]\,=\,$-$0.3 makes it clear
that HD~46703 has not experienced a third dredge-up. Also, the lack of an
overabundance of the s-process elements relative to the elements with the same
condensation temperature supports that conclusion. The [Zn/Fe-peak] and [S/Ti]
ratios are diagnostics for depletion, since they are elements with the same
chemical history, but with different condensation temperatures. For this
object, these element ratios are [Zn/Fe]\,=\,$+$0.82 and [S/Ti]\,=\,$+$1.04,
indicating that a very efficient depletion process took place.

\placefigure{fig_abund}

Our new results can be compared with the abundance analysis of this source published 
by \citet{luc84} more than twenty years ago. The spectra for their study date from 
1976, 1980 and 1981, and therefore reflect the status of the spectroscopic 
technology of that time: the 
signal-to-noise ratio of their spectra is rather low (as can be seen on their
Fig.~3).  While the authors derive abundances for a large number of elements,
the line-to-line scatter is fairly high (typically around 0.4\,dex, while ours is $\le$
0.2\,dex). Besides
the inferior quality of their data (compared to current standards), this large
scatter may also be attributed to the fact that their analysis is
based on spectra taken on different dates over several years (and hence different pulsational
phases), whereas our spectra were obtained over a shorter interval of seven days. 
Nevertheless, the abundances that are in common in both analyses are
in good agreement. Due to the large line-to-line scatter, however, the
depletion pattern is far less clear from the \citet{luc84} abundances. Their Zn abundance
is only moderately overabundant compared to iron ([Zn/Fe]\,=\,$+$0.2), and they
do not list an S abundance. The iron abundance they derive is
[Fe/H]\,=\,$-$1.56, $\sim$0.2\,dex higher than ours, which is due to
their slightly cooler model (6000\,K). Their model atmosphere parameters
(T$_{\rm eff}$\,=\,6000\,K, $\log g$\,=\,0.4, V$_t$\,=\,3.5\,km\,s$^{-1}$) are,
however, consistent with ours to within the errors. In a later paper
\citep{bon87}, the same authors derive nitrogen and sulphur abundances based on
a new spectrum and obtained [N/H]\,=\,$+$0.22 and [S/H]\,=\,$-$0.34;
their [N/H] value is similar to our value but their [S/H] value is higher than ours by +0.36. 

\citet{luc84} made a special remark on the H$\alpha$ line, which showed narrow
blue and red emission peaks that rise to the level of the continuum and that
vary slightly between their two spectra (their Figure~3).  The H$\alpha$ that we
observe is similar, with the blue and red emission components having equal strengths. 

\section{DISCUSSION}

HD~46703 has a high velocity (at galactic latitude +20$\arcdeg$), a low gravity,
and a low carbon abundance, all of which point to it being an evolved, low-mass, post-AGB star.  
The chemical depletion pattern indicates the presence of a circumbinary disk, for which there is also evidence from 
the observed intrinsic polarization \citep{tra94}. The depletion
of Fe and other Fe-peak elements make the estimate of the initial
metallicity difficult. Assuming the photospheric Zn abundance was
not affected by photospheric depletion, the initial metallicity was
about $-$1.0.

\citet{ruy06} present a systematic spectral energy distribution 
study of post-AGB stars around which disks are suspected.  This whole sample 
is  a significant fraction of all known post-AGB stars in the
Galaxy \citep[e.g.,][]{szc07}. Recently we found
similar objects in the Large Magellanic Cloud \citep[e.g.,][]{rey07b}.  
HD~46703 is among the minority of this disk sample that show no 
clear signature of a near-IR excess \citep[Figure A.1.]{ruy06}.
In most cases, dust is present at the sublimation temperature, which is not
the case in HD~46703. This suggests a larger disk radius.
Nonetheless, the disk is likely only resolvable
by interferometric means \citep{der07}.

HD~46703 is by no means the only disk post-AGB star for which the binary
orbit is determined \citep[e.g.,][and
references therein]{vanwin07}. One of the surprising results
of all the orbital studies so far is the fact that many objects survived
their phase of strong binary interaction in rather wide
orbits. Moreover, many systems, like HD~46703, are eccentric, despite the fact that
tidal effects as well as orbital-energy dissipation (during a likely
common envelope phase) are expected to rapidly circularize the orbits.
Episodic mass loss caused by periastron mass transfer in an elliptic
orbit has been proposed as a main mechanism \citep{sok00,bon08}, but also disk-binary
interactions may be strong enough to pump the eccentricity \citep{art91}.  
The latter mechanism is only efficient
when the dust extends outward to a distance of only a few times the major axis of the
orbit. For HD 46703, the lack of a near-IR excess points to the lack of
dust at the sublimation radius and thus a larger disk radius. 
Orbital pumping by the actual disk is not
likely the main eccentricity pumping mechanism for HD~46703.

 The global picture that emerges of all those objects is that a
binary star evolved in a system that is (presently) too small to accommodate a
full grown AGB star. During a badly understood phase of strong
interaction, a circumbinary dusty disk was formed, but the binary
system did not suffer a dramatic spiral in. What we observe now is an
F$-$K 
supergiant in a binary system that is surrounded by a
circumbinary dusty disk in a bound orbit. Understanding the evolution of the disks, 
as well as their impact on the evolution of the central stars, become
important challenges in the study of the final evolution of a significant fraction
of binary stars.

\acknowledgments 

We thank W. Zima for help with the period analysis using Period04, 
and we acknowledge the observing contribution of many undergraduate research students at 
Valparaiso University. 
BJH acknowledges financial support from the National Science Foundation
(9018032, 0407087), and MR acknowledges financial support from the Fund for Scientific Research - Flanders (Belgium).

\clearpage

\begin{deluxetable}{rrccrrccrrc}
\tablecaption{List of Radial Velocity Observations of HD~46703 \label{tab_rv}}
\tabletypesize{\footnotesize} \tablewidth{0pt} \tablehead{
\colhead{HJD}&\colhead{V$_{\rm r}$}&\colhead{Obs\tablenotemark{a}}&&\colhead{HJD}&\colhead{V$_{\rm r}$}&\colhead{Obs\tablenotemark{a}}&&\colhead{HJD}&\colhead{V$_{\rm r}$}&\colhead{Obs\tablenotemark{a}} \\
\colhead{}&\colhead{(km s$^{-1}$)}&\colhead{}&&\colhead{}&\colhead{(km s$^{-1}$)}&&\colhead{}&\colhead{}&\colhead{(km s$^{-1}$)}&\colhead{} }
\startdata
2443054.6670 &-106.20 & 1  && 2448663.3810 & -93.52 & 2 && 2449239.9700 & -96.14 &  3   \\
2443055.6670 &-105.00 & 1  && 2448714.7852 &  -88.53 &  3 && 2449244.0167 & -91.57 &  3  \\
2447757.6570 &-105.75 & 2 && 2448714.8364 &  -87.27 &  3 && 2449244.0291 & -92.37 &  3  \\
2447758.6500 &-100.31 & 2 && 2448719.7471 &  -84.59 &  3 && 2449264.0109 & -93.04 &  3   \\
2447760.6490 &-101.17 & 2 && 2448725.3480 & -84.27 &  2 && 2449287.0101 & -92.04 &  3  \\
2447761.6500 &-103.66 & 2 && 2448726.3280 & -84.24 &  2 && 2449384.7473 & -78.52 &  3   \\
2447945.4130 &-103.14 & 2 && 2448727.3420 & -85.20 &  2 && 2449405.4097 &  -77.9 &  4 \\
2447961.3080 &-102.22 & 2 && 2448728.3270 & -85.49 &  2 && 2449406.5792  & -77.9 & 4 \\
2447972.3380 & -98.28 & 2 && 2448730.3570 & -84.71 &  2 && 2449411.5368  & -77.0 & 4 \\
2447975.4080 & -99.81 & 2 && 2448731.3400 & -86.08 &  2 && 2449575.6330 & -107.11  & 2 \\
2447977.3510 &-100.92 & 2 && 2448733.7322 &  -86.07 &  3 && 2449576.6290 & -109.95   & 2 \\
2447996.3940 & -97.51 & 2 && 2448736.7225 &  -85.83 &  3  && 2449577.6350 & -108.66   & 2 \\
2447997.3380 & -97.49 & 2 && 2448756.7085 &  -83.32 &  3 && 2449580.6260 &-106.59   & 2 \\
2447998.3730 & -97.09 & 2 && 2448769.7557 &  -78.46 &  3 && 2449800.7381 & -98.24 &  3   \\
2448000.3580 & -99.62 & 2 && 2448778.7807 & -79.70 &  3  && 2449955.6550  & -82.23   & 2 \\
2448348.3360 & -97.39 & 2 && 2448875.9727 & -79.29 &  3  && 2450020.9430 & -75.68 &  3   \\
2448353.3460 & -98.88 & 2 && 2448922.8683 & -88.88 &  3 && 2450101.4540  & -83.96   & 2 \\
2448354.3560 &-100.51 & 2 && 2448952.8581 & -98.07 & 3 && 2450122.3620  & -86.99   & 2 \\
2448531.0476 &-105.83 & 3 && 2449084.3400 &-105.82 &  2 && 2450142.3580  & -99.70   & 2 \\
2448531.9238 &-107.71 & 3 && 2449086.3630 &-106.58 &  2 && 2450178.3600  & -100.56   & 2 \\ 
2448566.8739 & -99.52 & 3 && 2449089.3400 &-105.47 &  2 && 2450208.3400  & -104.77   & 2 \\  
2448600.8751 & -97.18 & 3  && 2449130.7615 &-104.27 & 3 && 2450232.3860  & -109.33   & 2 \\
2448627.8152 & -99.24 & 3  && 2449196.6340 &-100.20 &  2 && 2450293.6190  & -107.46   & 2 \\
2448628.8664 & -97.95 &  3 && 2449201.6230 &-103.17 &  2 && 2450295.6170  & -106.64   & 2 \\
\enddata
\tablenotetext{a}{References for the velocities: (1) \citet{luc84}, where we have estimated the fractional JD based on the observing data and location; (2) OHP$-$CORAVEL; (3) DAO$-$RVS; (4) derived from our high-resolution spectra used for the abundance analysis. }
\end{deluxetable}

\clearpage

\begin{deluxetable}{lcc}
\tablecaption{Orbital Solution of HD~46703\label{tab_orbit}}
\tabletypesize{\footnotesize} \tablewidth{0pt} \tablehead{
\colhead{Parameter (units)}& \colhead{Value}&\colhead{$\sigma$}  } \startdata
V$_o$  (km~s$^{-1}$) & $-$93.65 &  0.28  \\
K$_1$ (km~s$^{-1}$) & 16.85 & 0.54  \\
e &  0.323 &  0.023 \\
$\Omega$ &  1.194 &  0.090 \\
T$_o$ & 2448907.991 & 7.358  \\
Period (day) &   599.88 &   1.26 \\
$\sigma$ (km~s$^{-1}$) & 2.04   &\nodata   \\
a~sin{\it i} (AU) &0.879  &0.029  \\
f(M) (M$_\sun$) &0.252  &0.025  \\
\enddata
\end{deluxetable}

\clearpage

\begin{deluxetable}{rrrr}
\tablecolumns{4} \tablewidth{0pt} \tablecaption{Differential
Magnitudes of HD~46703 from VUO\tablenotemark{a}
\label{tab_difmag_vuo}} \tablehead{\colhead{HJD - 2,400,000} &
\colhead{$\Delta$V} & \colhead{HJD - 2,400,000} &
\colhead{$\Delta$R$_{\rm C}$}} \startdata
49,793.6586 & 1.030 & 49,951.9198  & 1.039 \\
49,802.6261 & 1.068 & 50,169.6203  & 1.306 \\
49,827.6171 & 1.062 & 50,184.6021  & 1.047 \\
49,840.5517 & 0.845 & 50,191.6014  & 1.064 \\
49,951.9178 & 0.873 & 50,385.7577  & 1.093 \\
49,965.8989 & 1.055 & 51,257.6332  & 1.137 \\
49,987.8408 & 0.903 & 51,882.7167  & 1.086 \\
50,001.8460 & 1.009 & 51,937.6837  & 1.087 \\ 
\enddata
\tablenotetext{a}{Table 3 is published in its entirety in the
electronic edition of the Astronomical Journal.  A portion is
shown here for guidance regarding form and content.}
\end{deluxetable}

\clearpage

\begin{deluxetable}{lrrrrrrr}
\tablecolumns{8} \tablewidth{0pt} \tablecaption{Magnitudes of 
HD~46703 on the Geneva Seven-Filter System\tablenotemark{a,b}
\label{tab_mag_geneva}} \tablehead{\colhead{HJD - 2,400,000} &
\colhead{U} & \colhead{B} & \colhead{V} & \colhead{B1} & \colhead{B2} & \colhead{V1} &
\colhead{G}} \startdata
46,742.605  &  10.652   & 8.753   & 9.067   & 9.735  & 10.136  &  9.799 &  10.117 \\
46,762.609  &  10.523   & 8.592   & 8.973   & 9.549  & 9.994  &  9.704  & 10.033  \\
46,765.555  &  10.502   & 8.555   & 8.943   & 9.506  & 9.961  &  9.676 &  10.005  \\
46,768.542  &  10.492   & 8.585   & 8.953   & 9.539  &  9.980 &   9.684 &  10.016  \\
46,775.473  &  10.514   & 8.630   & 9.000   & 9.577  & 10.027 &   9.739 &  10.059  \\
46,804.360  &  10.543   & 8.611   & 8.984   & 9.576  & 10.014  &  9.725 &  10.051  \\
46,814.350  &  10.661   & 8.736   & 9.077   & 9.719  & 10.153 &   9.838 &  10.164  \\
46,815.400  &  10.636   & 8.728   & 9.063   & 9.696  & 10.119  &  9.791 &  10.113  \\
\enddata
\tablenotetext{a}{Table 4 is published in its entirety in the
electronic edition of the Astronomical Journal.  A portion is
shown here for guidance regarding form and content.}
\tablenotetext{b}{Observations from 2446742$-$2450101
were made with the Swiss 0.7-m telescope and 
from 2452205$-$2453090 with the Flemish 1.2 Mercator telescope.}
\end{deluxetable}

\clearpage

\begin{deluxetable}{rrrrrr}
\tablecolumns{6} \tablewidth{0pt}
 \tablecaption{Measurements of the individual lines of each ion used in the abundance
analysis of HD 46703.\tablenotemark{a,b}
\label{line_ew}} \tablehead{\colhead{$\lambda$} &
\colhead{$\chi$} & \colhead{$\log\,gf$} & \colhead{W$_{\lambda}$} & \colhead{$\log\epsilon$} & \colhead{$\Delta\log\epsilon$}\\
\colhead{(\AA)} & \colhead{(eV)} &\colhead{} & \colhead{(m\AA)} & \colhead{} & \colhead{} } \startdata
\multicolumn{6}{c}{Abundance Results for C {\rm I}}  \\ \hline
   4228.33   &     7.68  &    -2.250   &     49   &       8.26   &       0.00\\
   4371.37   &     7.65  &    -1.970   &     80   &       8.31   &       0.05\\
   4770.03   &     7.46  &    -2.330   &     48   &       8.15   &      -0.11\\
   4771.74   &     7.46  &    -1.757   &    105   &       8.20   &      -0.06\\
   4775.90   &     7.49  &    -2.194   &     60   &       8.18   &      -0.08\\
   4817.37   &     7.48  &    -2.889   &     27   &       8.40   &       0.14\\
   5017.09   &     7.95  &    -2.431   &     21   &       8.21   &      -0.05\\
   5023.85   &     7.95  &    -2.196   &     22   &       8.00   &      -0.26\\
   5039.06   &     7.95  &    -1.775   &     94   &       8.53   &       0.27\\
   5040.13   &     7.95  &    -2.301   &     18   &       8.00   &      -0.26\\
\enddata
\tablenotetext{a}{Table~\ref{line_ew} is published in its entirety in the
electronic edition of the Astronomical Journal.  A portion is
shown here for guidance regarding form and content.}
\tablenotetext{b}{The following model atmosphere was adopted: 
T$_{\rm eff}$\,=\,6250\,K, $\log g$\,=\,1.0, $V_t$\,=\,3.0\,km\,s$^{-1}$,
and [M/H]\,=\,$-$1.5. }
\end{deluxetable}

\clearpage

\begin{deluxetable}{lrrrrrrr}
\tablecaption{Summary of the abundance results for HD 46703.\tablenotemark{a} \label{tab_abund}}
\tabletypesize{\footnotesize} \tablewidth{0pt}
\tablehead{\colhead{Ion} & \colhead{N} &\colhead{{\rule[0mm]{0mm}{4mm}$\overline{W_{\lambda}}$}}&\colhead{$\log\epsilon$}&\colhead{$\sigma_{\rm ltl}$}&\colhead{[X/H]}&\colhead{$\log\epsilon_{\odot}$ }&\colhead{T$_{\rm cond}$}\\
\colhead{} & \colhead{} &\colhead{(m\AA)} & \colhead{} & \colhead{} & \colhead{} & \colhead{} & \colhead{(K)}}
\startdata
C\,{\sc i  }  &  23 &  62 &  8.26 &  0.15 & $-$0.31 & 8.57 &   40\\
N\,{\sc i  }  &   5 &  46 &  8.16 &  0.14 & $+$0.17 & 7.99 &  123\\
O\,{\sc i  }  &   3 &  22 &  8.37 &  0.20 & $-$0.49 & 8.86 & \nodata \\
Na\,{\sc i  } &   2 &  21 &  4.96 &  0.13 & $-$1.37 & 6.33 &  958\\
Mg\,{\sc i  } &   5 &  49 &  6.05 &  0.22 & $-$1.49 & 7.54 & 1336\\
Al\,{\sc i  } &   2 & 155 &  4.45 &  0.14 & $-$2.02 & 6.47 & 1653\\
Si\,{\sc i  } &   1 &  11 &  6.10 & \nodata & $-$1.44 & 7.54 & 1310\\
Si\,{\sc ii } &   3 &  56 &  6.24 &  0.17 & $-$1.30 & 7.54 & 1310\\
S \,{\sc i  } &   4 &  22 &  6.63 &  0.15 & $-$0.70 & 7.33 &  664\\
Ca\,{\sc i  } &  18 &  37 &  4.78 &  0.18 & $-$1.58 & 6.36 & 1517\\
Sc\,{\sc ii } &   7 &  44 &  1.10 &  0.16 & $-$2.07 & 3.17 & 1659\\
Ti\,{\sc ii } &  31 &  49 &  3.23 &  0.16 & $-$1.79 & 5.02 & 1582\\
Cr\,{\sc i  } &   6 &  80 &  3.75 &  0.24 & $-$1.92 & 5.67 & 1296\\
Cr\,{\sc ii } &  16 &  56 &  4.11 &  0.18 & $-$1.56 & 5.67 & 1296\\
Mn\,{\sc i  } &   2 &  21 &  3.59 &  0.15 & $-$1.80 & 5.39 & 1158\\
Fe\,{\sc i  } & 106 &  44 &  5.78 &  0.16 & $-$1.73 & 7.51 & 1334\\
Fe\,{\sc ii } &  29 &  52 &  5.77 &  0.14 & $-$1.74 & 7.51 & 1334\\
Ni\,{\sc i  } &   6 &  22 &  4.67 &  0.18 & $-$1.58 & 6.25 & 1353\\
Zn\,{\sc i  } &   3 &  40 &  3.68 &  0.14 & $-$0.92 & 4.60 &  726\\
Y\,{\sc ii }  &   4 &  37 &  0.38 &  0.07 & $-$1.86 & 2.24 & 1659\\
Zr\,{\sc ii } &   2 &  16 &  0.94 &  0.14 & $-$1.66 & 2.60 & 1741\\
Ba\,{\sc ii } &   4 &  80 &  0.66 &  0.13 & $-$1.47 & 2.13 & 1455\\
\hline
\enddata
\tablenotetext{a}{The following model atmosphere was adopted: 
T$_{\rm eff}$\,=\,6250\,K, $\log g$\,=\,1.0, $V_t$\,=\,3.0\,km\,s$^{-1}$,
and [M/H]\,=\,$-$1.5. }
\end{deluxetable}

\clearpage

\clearpage

\clearpage

\clearpage

\clearpage

\clearpage

\end{document}